\documentclass[12pt]{article}
\usepackage{amssymb}
\usepackage{graphicx}
\usepackage{latexsym,cite}
\newcommand{\eq}{\begin{equation}}
\newcommand{\en}{\end{equation}}
\newcommand{\be}{\begin{equation}}
\newcommand{\ee}{\end{equation}}
\newcommand{\qe}{\end{equation}}
\newcommand{\ear}{\begin{eqnarray}}
\newcommand{\eqa}{\begin{eqnarray}}
\newcommand{\ba}{\begin{eqnarray}}
\newcommand{\ea}{\begin{eqnarray}}
\newcommand{\rae}{\end{eqnarray}}
\newcommand{\ena}{\end{eqnarray}}
\newcommand{\beq}{\begin{equation}}
\newcommand{\eeq}{\end{equation}}
\newcommand{\bea}{\begin{eqnarray}}
\newcommand{\eea}{\end{eqnarray}}
\newcommand{\Z}{\mathbb{Z}}
\newcommand{\N}{\mathbb{N}}
\newcommand{\bra}{\langle}
\newcommand{\ket}{\rangle}

\begin{document}
\begin{titlepage}
\vskip0.5cm
\begin{flushright}
DFTT 27/05\\
IFUP-TH 2005-30\\
DIAS-STP-05-11 \\
\end{flushright}
\vskip0.5cm
\begin{center}
{\Large\bf
On the effective string spectrum of the tridimensional $\Z_2$ gauge model}
\end{center}
\vskip1.3cm
\centerline{
M. Caselle$^{a}$, M. Hasenbusch$^{b}$
 and M. Panero$^{c}$}
 \vskip1.0cm
 \centerline{\sl  $^a$ Dipartimento di Fisica
 Teorica dell'Universit\`a di Torino and I.N.F.N.,}
 \centerline{\sl Via Pietro~Giuria~1, I-10125 Torino, Italy}
 \centerline{\sl
e--mail: \hskip 1cm
 caselle@to.infn.it}
 \vskip0.4 cm
 \centerline{\sl  $^b$  Dipartimento di Fisica dell'Universit\`a di Pisa
 and I.N.F.N.,}
   \centerline{\sl Largo Bruno~Pontecorvo~3, 
                  I-56127 Pisa,
                        Italy}
 \centerline{\sl
e--mail: \hskip 1cm
 Martin.Hasenbusch@df.unipi.it}
\vskip0.4 cm
 \centerline{\sl  $^c$ School of Theoretical Physics,
Dublin Institute for Advanced Studies,}
 \centerline{\sl
                  10~Burlington Road, Dublin~4,
                              Ireland}
 \centerline{\sl
e--mail: \hskip 1cm
 panero@stp.dias.ie}
 \vskip1.0cm

\begin{abstract}
We study the $\Z_2$ lattice gauge theory in three dimensions, and present 
high precision estimates for the first few energy levels of the string 
spectrum. These results are
obtained from new numerical data for the two-point Polyakov loop 
correlation function, which is measured in the 3d Ising spin system 
using duality.
This allows us to perform a stringent comparison with the predictions
of effective string models.
We find a remarkable agreement between the numerical estimates
and the Nambu-Goto predictions for the  energy gaps at intermediate and
large distances. The precision of our data
allows  to distinguish clearly between the predictions of the 
full Nambu-Goto action and the simple free string model up to an 
interquark distance $r \approx 10/\sqrt{\sigma}$.
At the same time, our results also
confirm the breakdown of the effective picture at short distances, 
supporting the hypothesis that terms which are not taken into account 
in the usual Nambu-Goto string formulation yield a non-trivial shift to 
the energy levels.
Furthermore, we discuss the theoretical implications of these results.
\end{abstract}
\end{titlepage}

\setcounter{footnote}{0}
\def\thefootnote{\arabic{footnote}}

\section{Introduction}
\label{introsect}
Understanding in detail the mechanisms which 
govern
the behaviour of physical observables in the confinement phase of quantum 
gauge theories is still a challenge.  
Even the simplest confined system --- a heavy quark-antiquark pair in a pure 
gauge model without dynamical matter
fields --- is a setting where a number of interesting aspects become manifest;
the potential associated with such a system is asymptotically 
linearly rising
at large interquark distances, whereas at finite distance it develops 
non-trivial corrections, which are expected to be predicted by some kind 
of effective theory.

The idea that the confining flux lines joining the two colour sources get 
squeezed in a thin tube 
explains that at asymptotically large distances the 
potential energy of the system is proportional to the 
quark-antiquark distance $r$. 
Assuming that the tube vibrates along the transverse
directions, one can derive effective string corrections
affecting the potential $V(r)$ for finite values of $r$.
This infrared picture breaks down as $r$ approaches $1/m$ from above,
where $m$ is the mass of the lightest glueball. At these small distances, 
glueball radiation has to be taken into account, leading to effects that 
cannot be described by the effective string model and which are specific
to the given gauge model.

The theoretical background of the effective string in this setting is 
well-known: the pioneering works by 
L\"uscher, M\"unster, Symanzik and Weisz \cite{Luscher:1980fr,Luscher:1980iy,Luscher:1980ac} date back to 
the Eighties, and over the last decades considerable progress has been achieved in understanding the details 
of the picture \cite{Alvarez:1981kc,Pisarski:1982cn,Dietz:1982uc,Arvis:1983fp,Ambjorn:1984yu,Olesen:1985pv,Olesen:1985ej,Polchinski:1991ax,Luscher:2004ib}.

These theoretical predictions can be confronted with data obtained 
from lattice gauge theory:  
One can check the extent to which the theoretical expectations 
actually match the results of Monte Carlo simulations for a wide range of
parameters and for different types of gauge theories,
in particular simpler prototype models for confinement than QCD
\cite{Caselle:2002rm,Caselle:2002ah,Caselle:2004jq,Caselle:2004er,Caselle:2005xy,Allen:1998wp,Juge:2002br,Juge:2003sz,Juge:2003ge,Juge:2004xr,Majumdar:2004qx}.

The aim of this work is to study numerically the fine details in the behaviour of the excitation energies $E_n(r)$ which contribute to the partition function $Z(r,L)$ describing the confined quark-antiquark ($Q\bar{Q}$) system in Euclidean space-time with a compactified direction of size $L$:
\eq
\label{partitionfunction}
Z_{Q\bar{Q}}(r,L)= \sum_{n=0}^{\infty} w_n e^{-E_n(r)L} 
\en
where $w_n$ is the multiplicity factor associated with the energy level $E_n$.
Neglecting glueball 
radiation,\footnote{At large $r$, the glueball threshold is indeed far from the lowest-lying energy states, whereas at short distances this is no longer true --- see subsection~\ref{excitedlevelsubsect} and table~\ref{lightstatestab}.}
such a partition function is proportional to the two-point correlation 
function among Polyakov lines $G(r)$, 
which can be evaluated in numerical simulations.

Here, we focus our attention onto the results obtained from the confining
phase of $\Z_2$ lattice gauge  theory in three dimensions:
a model whose properties (remarkably, its duality with respect to the Ising 
spin model) are well-known, 
and allow to reach high precision in the determination of the 
interquark potential, 
{\sl even for large values of the interquark distance $r$,
or of the lattice size in the 
time-like direction $L$}. 
This in turn allows very stringent comparisons with the
existing theoretical descriptions and in particular with the 
effective string models which we shall discuss
below. In particular in this paper we will concentrate on the 
study of the behaviour of the 
first few energy levels $E_n$ as a function of the interquark distance 
$r$. We will compare our numerical data with the predictions 
of both the Nambu-Goto (NG) model and its free string limit. Although
these two models give very similar results, the
precision of our numerical simulation is 
sufficient to distinguish between them. 
We shall see that while the lowest state $E_0$ shows
deviations from the effective string prediction even for 
rather large distances (as already pointed out in
several papers) the energy gap $E_1-E_0$ clearly approaches the prediction
of the Nambu-Goto effective string model as the distance increases.
This suggests 
that the  breakdown of the effective string model is
due to additional contributions at  short distances which 
lead to an overall shift on the effective string spectrum. In addition to 
gauge model specific effects, such as glueball radiation, such 
a contribution might be due to the so-called Liouville mode,
which is neglected in deriving the effective string model predictions from
the original reparametrisation-invariant string action.

Let us finally mention that other studies of the excited states of the spectrum 
for a static quark-antiquark pair have been done in recent years,
both in the Ising model and in other LGT's 
\cite{Allen:1998wp,Juge:2002br,Juge:2003sz,Juge:2003ge,Juge:2004xr,Majumdar:2004qx}, but they used different numerical techniques and 
did not reach distances large enough to appreciate 
the effects that we discuss in this paper. The novelty of our approach is that,
by suitably combining the results of our simulations for all 
intermediate values or the interquark distance, we were able to directly 
evaluate the partition function 
of the model eq.~(\ref{partitionfunction}). From it, by fitting its functional
dependence on the energy levels --- see the right-hand side of 
eq.~(\ref{partitionfunction}) --- for different values of $r$ and $L$ we could 
extract high precision estimates of the first  few energy levels.

This paper is organised as follows.
In section~\ref{theorysect} we will recall some basic theoretical ideas 
underlying the effective description of the interquark potential, 
focusing our attention onto the bosonic string models that are expected 
to mimic its large distance behaviour; in particular, we will discuss 
the Nambu-Goto string and its free string limit. In 
section~\ref{algorithmsect}, after reviewing  the basic properties of the 
$\Z_2$ lattice gauge theory in $D=3$, we will present the features of our 
simulation algorithm. Then, in section~\ref{resultssect} we will show the 
numerical results; the latter will be discussed in 
section~\ref{conclusionssect}, in a comparison with similar studies; 
finally, we will conclude with a few remarks.

\section{Theory}
\label{theorysect}

A quantitative description of the $Q\bar{Q}$ potential in confining gauge theories --- see \cite{Bali:2004tm} and references therein for a review --- can be formulated in terms of effective models, which are expected to provide phenomenologically consistent predictions for the behaviour of $V(r)$ in the regime of interquark distances typical of the hadronic world.

At large enough interquark distances, the 
region of the perturbative vacuum among two heavy sources
gets ``stretched'' into a vortex-like configuration, whose excitations 
correspond to string-like modes: this 
is the scenario underlying the renowned effective string picture
for confinement in the IR \cite{Luscher:1980fr,Luscher:1980iy,Luscher:1980ac}, 
based on the idea that the flux lines among the sources in a pure gauge theory
get squeezed in a thin, almost uni-dimensional tube. As a consequence, 
the asymptotic behaviour of the potential is a linear rise. Remarkably enough,
according to this picture, in principle one can obtain physical information
about the confined system without knowing the dynamics of the microscopic 
degrees of freedom and the details of the gauge group. 

In the following of this paper, 
we focus on sufficiently large interquark distances,
where the effective string picture holds, and try to identify the string 
model which yields predictions that match best with our numerical results 
for the energy spectrum.

At finite distances, the leading correction to the linear 
interquark potential predicted by the effective string model 
is a Casimir effect due to the (harmonic) oscillations that can 
set in the finite-size string with fixed ends: 
they induce a $1/r$ contribution to the IR interquark potential:
\eq
\label{potenziale}
V(r)=\sigma r + \mu - \frac{\pi (D-2)}{24r} + O\left( \frac{1}{r^2} \right)
\en
where $D$ is the number of space-time dimensions, $\sigma$ is the string 
tension, and $\mu$ is a constant. The $1/r$ term in eq.~(\ref{potenziale}) 
is known as the ``L\"uscher term'': it can be predicted under the assumption 
that, at leading order, the string fluctuations along the $D-2$ transverse
 directions are described by free, massless boson fields, and its numerical 
coefficient can also be obtained via CFT arguments. 

According to this ``free string'' picture, the excitation spectrum is expected
to be simply described as a tower of equally-spaced levels labelled by a 
non-negative integer $n$:
\eq
\label{spettrooscarm}
E_n=E_0 + \frac{\pi}{r}n 
\en
and, in general, degeneracies are expected for $n \ge 1$.

The prediction for $V(r)$ in eq.~(\ref{potenziale}) can be refined,
 making more precise assumptions about the dynamics describing string
 vibrations: more precisely, the partition function $Z(r,L)$ associated 
to the $Q\bar{Q}$ sector of the gauge theory (which approximates the 
Polyakov
loop two-point correlation function $G(r)$) can be expressed as a string
partition function:
\eq
\label{prp0conazeff}
G(r) = \bra P^\dagger (r) P(0) \ket = \int \left[ \mathcal{D} h \right] e^{-S_{\mbox{\tiny{eff}}}} 
\en
where $S_{\mbox{\tiny{eff}}}$ denotes the effective action for the world sheet
spanned by the string. In eq.~(\ref{prp0conazeff}), the functional integration
 is done over world sheet configurations which have fixed boundary conditions
 along the space-like direction, and periodic boundary conditions along the 
compactified, time-like direction --- the Polyakov lines being the fixed 
boundary of the string world sheet. In general, the effective string action 
$S_{\mbox{\tiny{eff}}}$ describing string dynamics also encodes string 
interactions.

A particular choice for $S_{\mbox{\tiny{eff}}}$ is to assume that it is 
proportional to the area spanned by the string world sheet:
\eq
\label{azionenambugoto}
S_{\mbox{\tiny{eff}}}= \sigma \cdot \int d^2 \xi \sqrt{ \det g_{\alpha \beta} }
\en
which is the Nambu-Goto action \cite{Nambu:1970,Goto:1971ce,Nambu:1974zg}. 

The Nambu-Goto model can be quantised along two almost equivalent ways: 
\begin{description}
\item{a]}
Fix the reparametrisation and Weyl invariance of
eq.~(\ref{azionenambugoto}) using the so-called \emph{physical gauge} 
(see for instance~\cite{Alvarez:1981kc,Arvis:1983fp}) in which the 
longitudinal degrees of freedom of the string
are identified with the coordinates of the plane in which the two 
Polyakov loops lie. This gauge choice is anomalous 
(unless 
$D=26$)~\cite{Goddard:1973qh}: the anomaly manifests itself as a breaking of 
Lorentz invariance, but it can be shown that the anomalous contribution is a 
rapidly decreasing function of the interquark distance~\cite{Olesen:1985pv}.
The effective string model is the quantum field theory which one obtains by
simply neglecting this anomaly. The obvious implication of this assumption is
that the effective string theory predictions are bound to hold only  for large
enough interquark distances. The critical distance below which the picture is
no longer valid cannot be deduced from the theory. It can only be obtained
numerically by comparing prediction and simulations. In the limit in which 
the anomaly can be neglected 
it is possible to proceed to formal quantisation of the resulting 
action~\cite{Alvarez:1981kc,Arvis:1983fp} --- which has now been reduced 
to an ordinary QFT of the transverse degrees of freedom only, interacting
via a square-root-type potential. The spectrum obtained in this way is: 
\eq
\label{enarvis}
E_n(r)= \sigma r \sqrt{ 1 + \frac{2 \pi}{ \sigma r^2} \left( n - \frac{D-2}{24} \right)  } \;\; ,  \;\;\;\; n \in \N
\en
and the $w_n$'s --- \emph{i.e.} the multiplicity factors which appear in eq.~(\ref{partitionfunction}) --- are given by:
\eq
\label{wnfactors}
w_n= \sum^{n}_{i_1=0} \dots \sum^{n}_{i_{D-2}=0}
\left[ P\left(i_1 \right) \dots P\left(i_{D-2} \right) \delta_{n, i_1 + \dots + i_{D-2} }\right]
\en
where $P(i)$ denotes the number of partitions of $i$. In particular, in three
space-time dimensions, the $w_n$ coefficients are simply the partitions of $n$,
and we end up with the following expression for the effective string partition
function:
\eq
\label{partitionfunctionasaseries3d}
Z_{Q\bar{Q}}(r,L) = \sum_{n=0}^{\infty} P(n) e^{- E_n(r) L}
\en

\item{b]}
Alternatively, one could use the reparametrisation invariance to reach the
conformal gauge (where the NG action is equivalent to the free
string action) and then quantize it using the so-called covariant  
quantisation~\cite{gsw,polbook} in the background of
two ``zero branes'', which play the role of the Polyakov loops. 
In this framework, the anomaly shows up with the appearance of  an
additional field (the so-called ``Liouville mode''). 
If one assumes, as above, that at large distance this mode
can be neglected, then one can quantize the string as it is usually done 
for the critical bosonic string, and re-obtain all the 
previous results~\cite{Billo:2005iv}. In particular, one exactly finds
the Nambu-Goto spectrum of eq.~(\ref{enarvis}). The major advantage of 
the latter procedure is that it makes the role of the Liouville mode explicit,
and it could in principle offer a clue to guess its contribution to the 
effective string prediction at shorter interquark distances.

\end{description}
One should stress that the Nambu-Goto action is by no reason the only 
possible choice for the effective string  action. It is the simplest
choice and has a nice geometrical interpretation.
Thus in these last years much effort has been devoted to the construction of
alternative string actions. Among these a particularly interesting proposal 
appeared a few years ago in~\cite{Polchinski:1991ax}. We shall not further 
discuss this proposal in the present paper, we only mention here that it can
be shown that the first two perturbative orders in the string tension expansion
of this effective string coincide with the Nambu-Goto 
ones~\cite{Drummond:2004yp} and thus, at least at low temperature and large 
distance, the two effective strings should follow the same behaviour.

\section{General setting and the algorithm}
\label{algorithmsect}

In this paper, we restrict our attention to the 
$\Z_2$ lattice gauge theory in $D=3$:
 the dynamics of the $U_\mu (x)$ bond variables (taking values in $\Z_2$) 
is expressed by the standard Wilson action:
\eq
\label{wilsonaction}
S= - \beta \sum_{\Box} U_{\Box} \;\;\; , \;\;\; U_{\Box}= \prod_{l \in \partial \Box} U_l
\en

For values of the coupling $\beta$ below a critical value 
 $\beta_c =0.76141346(6) $ \cite{Deng:2003wv},
the system is in the confinement phase, whereas for $\beta > \beta_c$ 
the system is deconfined; the deconfinement transition at $\beta=\beta_c$ 
is a second-order one. This model also possesses an (infinite-order) 
``roughening transition'' at $\beta_r = 0.47542(1)$ \cite{Hasenbusch:1996eu}
(in the confined phase), which separates the strong coupling regime 
(for $\beta< \beta_r $) from the so-called ``rough phase'' 
(for $\beta_r < \beta < \beta_c $).

This model is related to the Ising spin model in $D=3$ by an exact duality 
mapping \emph{\`a la} Kramers and Wannier: a $\Z_2$ Fourier transform on 
the plaquette variables $U_{\Box}$ maps the partition function of the gauge
model to the partition function of the spin system, evaluated at a different
coupling:
\eq
\label{duality}
Z_{\mbox{\tiny{gauge}}} (\beta) \propto  Z_{\mbox{\tiny{spin}}} (\tilde{\beta}) \;\;\; , \;\;\; \mbox{with: $\tilde{\beta}=-\frac{1}{2}\log\left[\tanh(\beta)\right]$}
\en
and analogous relations hold for other observables; in particular, it is 
interesting to see that the $\Z_2$ gauge theory glueballs can be interpreted 
in terms of bound states of the fundamental scalar in the spin model 
\cite{Caselle:2001im}.

In the case of our interest, adding external source terms for the gauge model
(for instance, a pair of Polyakov loops) amounts to 
introducing sets of topological defects in the spin system. As a result, 
the partition function associated with the $Q\bar{Q}$ gauge system on the 
left-hand side of eq.~(\ref{partitionfunction}) is proportional to the 
partition function of the spin system with anti-ferromagnetic coupling on 
a set of links, namely:
\eq
\label{zspindefects}
Z_{\mbox{\tiny{spin}},Q\bar{Q}}(r,L) = \sum_{ \{ s_i \} } 
\exp \left( \tilde{\beta} \sum_{\bra i,j \ket} J_{\bra i,j \ket} s_i s_j \right)
\en
where $\sum_{ \{ s_i \} }$ denotes the sum over spin configurations, the $i$ 
and $j$ indices denote lattice sites, and the $s_i$ spin variables interact 
with their nearest-neighbours only. The value of the $J_{\bra i,j \ket}$ 
coupling is $+1$ everywhere, except on a set of bonds, which pierce a surface 
(in the direct lattice) having the source worldlines as its boundary: 
for such a set of bonds, $J_{\bra i,j \ket}=-1$.
The Polyakov-loop correlation function is then given by
\eq
\label{Gspin}
G(r) = \frac{Z_{\mbox{\tiny{spin}},Q\bar{Q}}(r,L)}{Z_{\mbox{\tiny{spin}},Q\bar{Q}}(0,L)}
\en
Our numerical algorithm (see also 
\cite{Hasenbusch:1993pz,Caselle:1994df,Hasenbusch:1996eu,Caselle:2002ah} 
for further details) exploits this duality of the model, simulating the Ising 
spin system, and measuring ratios of the partition functions associated with 
different stacks of defects --- which can be used to express the expectation 
values of Polyakov loops pairs in the original gauge model. 

The $\Z_2$ spin variables are stored with multi-spin coding implementation,
and updated by means of a microcanonical demon-update, in combination with a
canonical update of the demon \cite{Rummukainen:1992fc}. In order to reduce 
the statistical errors, we use the snake-algorithm method 
\cite{deForcrand:2000fi,deForcrand:2000qn,Pepe:2001cx} furtherly improved by 
a hierarchical organisation of sublattice updates; this results in an 
efficient algorithm which allows to reach a high degree of precision, 
compared to direct numerical simulations in the standard setting of the 
theory (see, for instance, \cite{Caselle:2002rm}; see also 
\cite{Caselle:1996ii} for a comparison of results from simulations in the 
direct setting and in the dual setting).

A particularly useful advantage of numerical simulations in the dual setting 
is the fact that this method overcomes the problem of exponential 
signal-to-noise ratio decay, which is usually found when studying the 
interquark potential $V(r)$ at larger and larger distances (see also 
\cite{Panero:2005iu}, where the technique was applied to compact QED).

\section{Numerical results}
\label{resultssect}
We have performed a set of new simulations at $\tilde \beta=0.236025$, 
$0.24607$ and  $0.27604$, where the finite temperature phase transition occurs 
at $L=4,3$ and $2$, respectively \cite{Caselle:1995wn}. 
The correlation length at these values of
$\tilde \beta$ is $\xi = 1.456(3),1.040(2),0.644(1)$,
respectively. 
These estimates are obtained by inter- and extrapolation of Monte Carlo results
for $\xi$ given in~\cite{Caselle:1997hs,Caselle:1999tm} and the analysis of the low
temperature series~\cite{Arisue:1994wc}. We should keep these estimates in mind, since we can only expect to see a string spectrum described by some effective string theory as long as $E_n - E_0 < m$, where $m$ is the lightest glueball mass (or the inverse correlation length in the Ising spin model).

We have chosen these rather large values of $\tilde \beta$ mainly for 
technical reason. Since we use a local update algorithm, the effort required for the 
simulation at a given statistical accuracy grows like 
$\xi^{d+z} \approx \xi^{5}$. Hence for a given amount of CPU-time much more 
accurate results can be obtained staying at a moderate correlation length.

A crucial question is to understand how much our results are affected by scaling
corrections. The basic assumption entering the Nambu-Goto string action is 
that the rotational and translational symmetries of the continuum are restored. Leading scaling corrections which are proportional to $\xi^{-\omega}$ with $\omega = 0.821(5)$ \cite{Deng:2003wv} are not related with the breaking of these symmetries. The breaking of the symmetries comes with a larger exponent $\rho \approx 2$ \cite{Campostrini:1997sn}. Hence we might expect to see the proper string spectrum at values of $\tilde \beta$, where otherwise scaling corrections are still large. 

We have simulated lattices of the size $64^2 \times L$, $48^2 \times L$   
and $32^2 \times L$ at $\tilde \beta=0.236025$, $0.24607$ and  $0.27604$,
respectively. In order to get a numerical result for $G(r)$ in a range 
$0 \le r \le r_{\mbox{\tiny{max}}}$ we have computed, in contrast to our previous work, 
$G(r+1)/G(r)$ for all $0 \le r < r_{\mbox{\tiny{max}}}$. 
Also the statistics, in particular
for small values of $L$ is considerably (\emph{i.e.} about a factor 10) larger 
than in our previous work. 

In a first step, we analyse the ratio $G(r+1)/G(r)$ itself. 
To this end, we define an effective string tension $\sigma(r,L)$
as the solution of
\begin{equation}
 \frac{G(r+1)}{G(r)} = \frac{Z_{Q\bar{Q}}(r+1,L)}{Z_{Q\bar{Q}}(r,L)}
\end{equation}
with respect to $\sigma$, where $G(r+1)/G(r)$ is the numerical result
of our simulation and $Z_{Q\bar{Q}}(r,L)$ the theoretical 
prediction~(\ref{partitionfunctionasaseries3d}) with either
the free string energy levels of eq.~(\ref{spettrooscarm}) or those derived
from the Nambu-Goto action, in eq.~(\ref{enarvis}).

\begin{figure}
\begin{center}
\includegraphics[width=12.3cm]{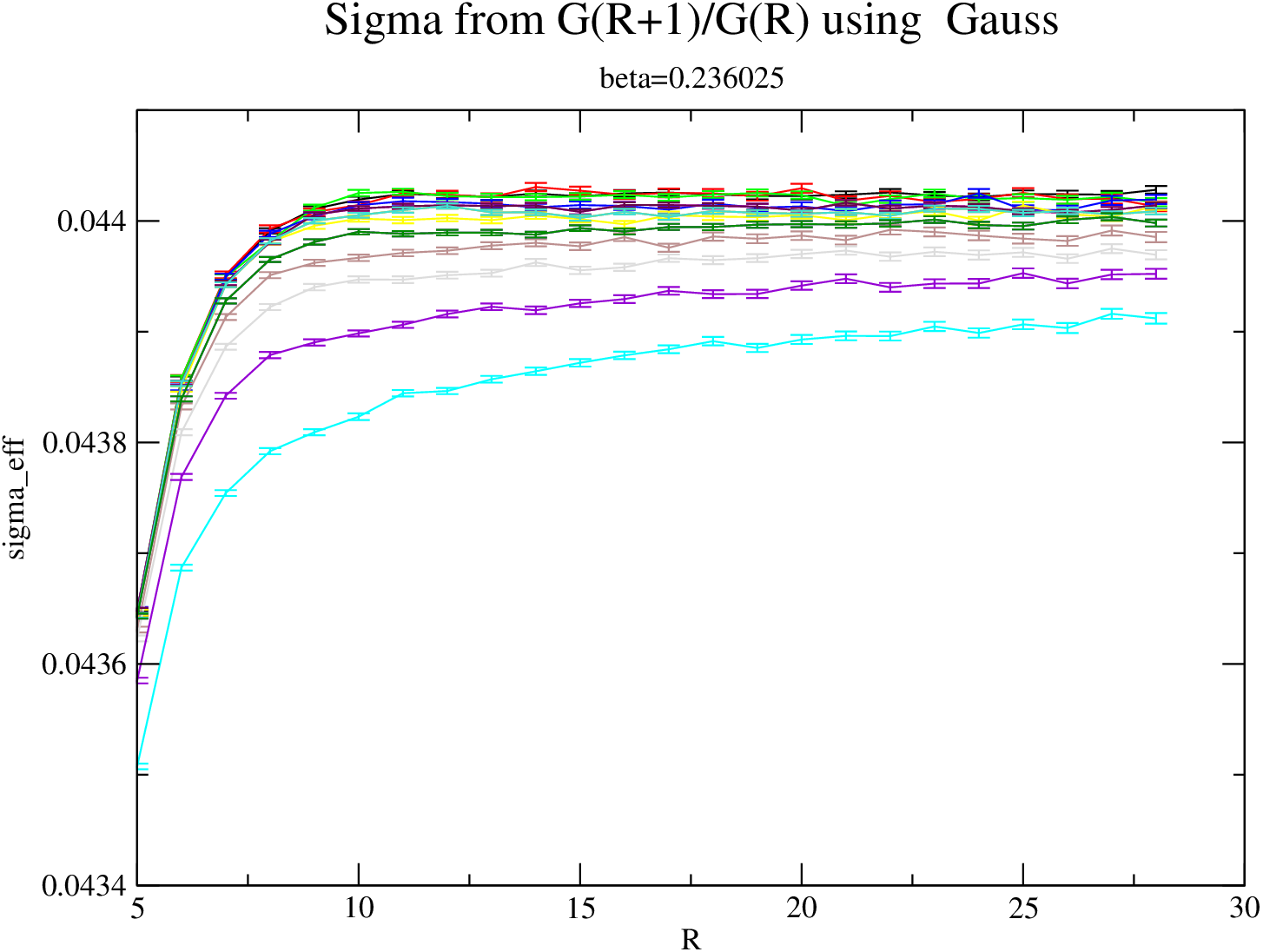}
\vskip0.5cm
\includegraphics[width=12.3cm]{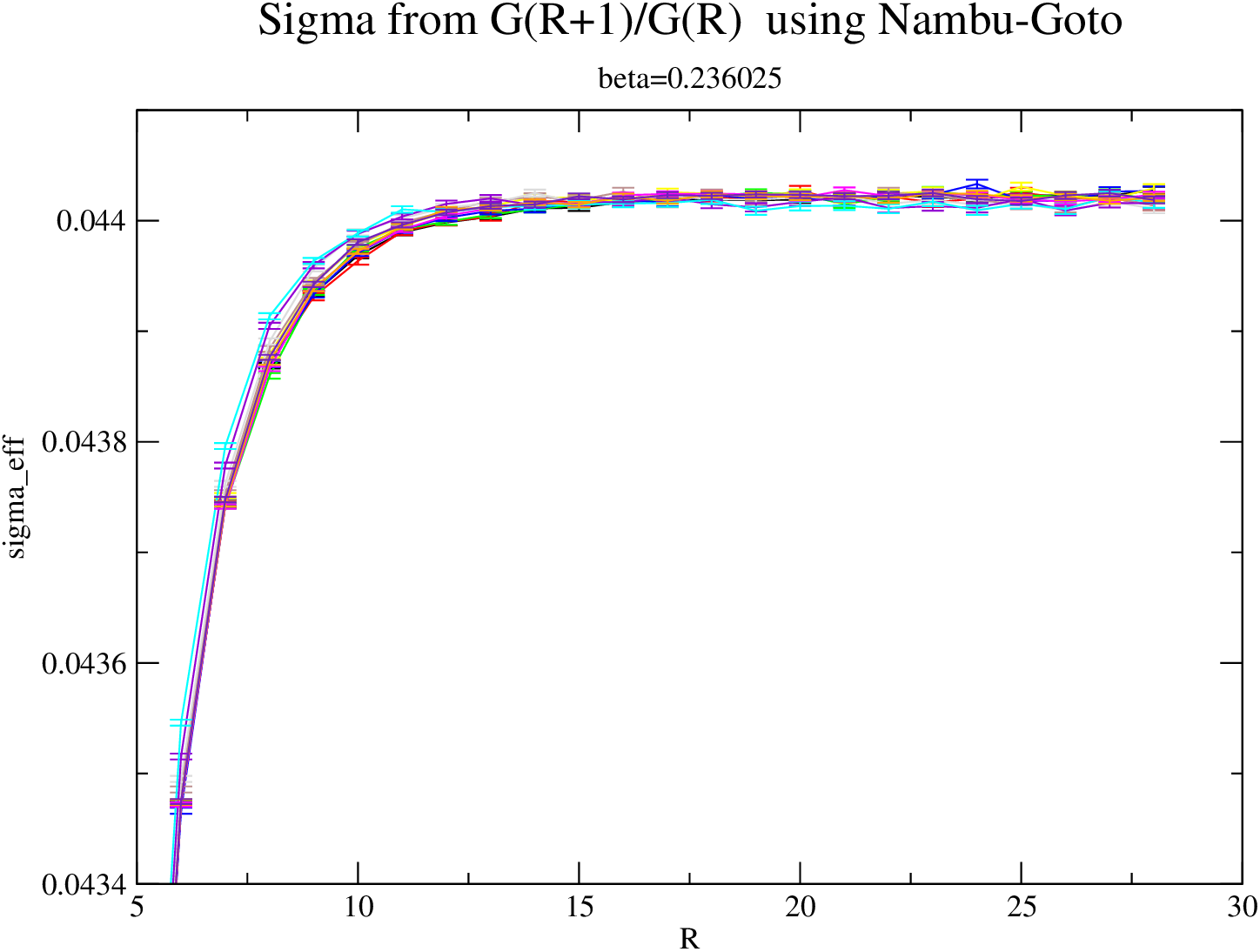}
\end{center}
\caption{ 
\label{Lmany4}
Effective string tension computed from $G(r+1)/G(r)$ using
eq.~(\ref{partitionfunctionasaseries3d}). For $\beta=0.236025$, 
$L=80,50,40,30,28,26,24,22,20,18,16,14$. 
Upper plot:  free field energy levels; lower plot: Nambu-Goto energy levels.
}
\end{figure}

\begin{figure}
\begin{center}
\includegraphics[width=12.3cm]{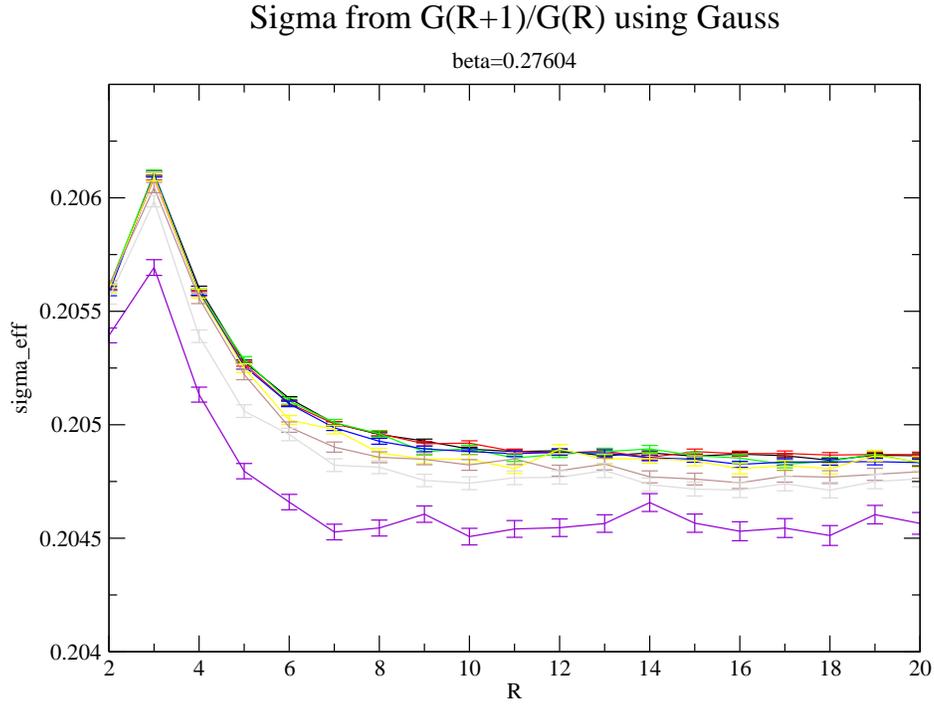}
\vskip0.5cm
\includegraphics[width=12.3cm]{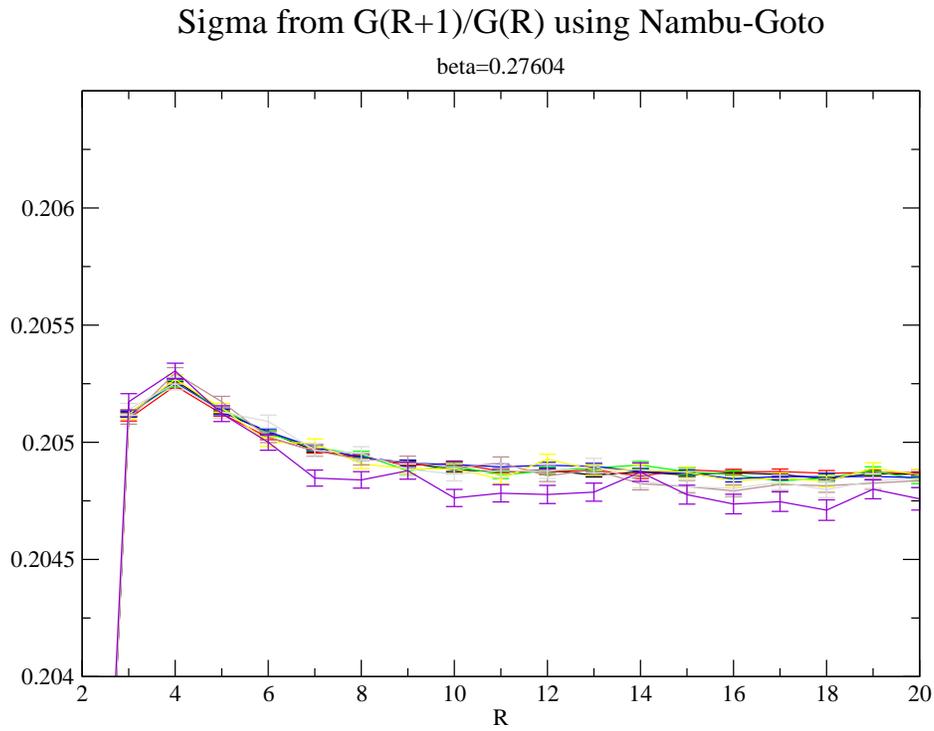}
\end{center}
\caption{
\label{Lmany2}
Effective string tension computed from $G(r+1)/G(r)$ using
eq.~(\ref{partitionfunctionasaseries3d}) For $\beta=0.27604$,
$L=40,30,20,16,14,12,10,8$.
Upper plot:  free field energy levels; lower plot: Nambu-Goto energy levels.
}
\end{figure}

In figures~\ref{Lmany4}, \ref{Lmany3} and \ref{Lmany2} 
we have plotted our results for $\tilde \beta=0.236025$, $0.24607$ and
$0.27604$, respectively.

Let us first have a closer look at 
figure~\ref{Lmany4},
where we have plotted the effective $\sigma$ for 
$\tilde \beta=0.236025$.
The upper figure shows the results
from free field energy levels of eq.~(\ref{spettrooscarm})
while the lower one uses the Nambu-Goto
energy levels in eq.~(\ref{enarvis}).
In the free field case we see 
quite a big spread of the
curves for different values of $L$.
On the other hand, for the Nambu-Goto energy levels, the curves
obtained for different $L$ fall nicely on top of each other.  Starting
from about $r=14$ the results from all values of $L$ are compatible within
error-bars. Also for smaller $r$ the spread among different values of $L$
is much smaller than in the free field case.

Looking at the $r$-dependence of $\sigma_{\mbox{\tiny{eff}}}$ for $L=80$ the situation 
is quite different. Within error
bars, the  $\sigma_{\mbox{\tiny{eff}}}$ is constant starting from $r=11$  
in the free field case. On the other hand, with the Nambu-Goto ansatz we still
see deviations from the large $r$ limit 
up to $r=16$. 
Averaging the $\sigma_{\mbox{\tiny{eff}}}$ from the free field ansatz 
for $L=80$ and $r>24$ we get $\sigma=0.0440244(15)$. To estimate possible
systematic errors we compare this estimate 
with the result from the Nambu-Goto ansatz
(0.0440232(15)) and the corresponding results for $L=50$: 
$\sigma=0.0440196(19)$ and $0.0440201(19)$ 
using the free string and the Nambu-Goto ansatz, respectively.
As our final estimate, which is compatible with all the results given above, 
we quote $\sigma=0.044023(3)$. 

Looking at figures~\ref{Lmany3} and~\ref{Lmany2} we see that 
also for $\tilde \beta=0.24607$ and $\tilde \beta=0.27604$ the curves 
for different $L$ fall nicely on top of each other in the case of the 
Nambu-Goto ansatz, while there is a clear spread in the case of the 
free field ansatz.  Finally, we notice that the $r$-dependence of 
$\sigma_{\mbox{\tiny{eff}}}$ is quite different for $\tilde \beta=0.27604$ compared with 
$\tilde \beta=0.24607$ and $\tilde \beta=0.236025$, \emph{i.e.} there are sizable 
scaling corrections. 
From $L=60$ for $\tilde \beta=0.24607$ and $L=40$ for $\tilde \beta=0.27604$
and large values of $r$ we get $\sigma=0.082520(3)$ and $\sigma=0.20486(1)$ 
as our final 
estimate of the string tension at $\tilde \beta=0.24607$ and 
$\tilde \beta=0.27604$, respectively. The error quoted should include
systematic errors. 
We get for the dimensionless combination $\sigma \xi^2 = 0.0933(4)$, $0.0893(4)$
and $0.0850(3)$
for $\tilde \beta=0.236025$, $0.24607$ and  $0.27604$, respectively.
The universal limit is given by 
$\lim_{\tilde \beta \rightarrow \tilde \beta_c} \sigma \xi^2=0.1056(19)$ 
\cite{Agostini:1996xy}. 
\begin{figure}
\begin{center}
\includegraphics[width=12.3cm]{sigma3g.eps}
\vskip0.5cm
\includegraphics[width=12.3cm]{sigma3ng.eps}
\end{center}
\caption{
\label{Lmany3}
Effective string tension computed from $G(r+1)/G(r)$ using
eq.~(\ref{partitionfunctionasaseries3d}) For $\tilde \beta=0.24607$,
$L=60,30,24,22,20,18,16,14$.
Upper plot:  free field energy levels; lower plot: Nambu-Goto energy levels.
}
\end{figure}

\begin{figure}
\begin{center}
\includegraphics[width=14.3cm]{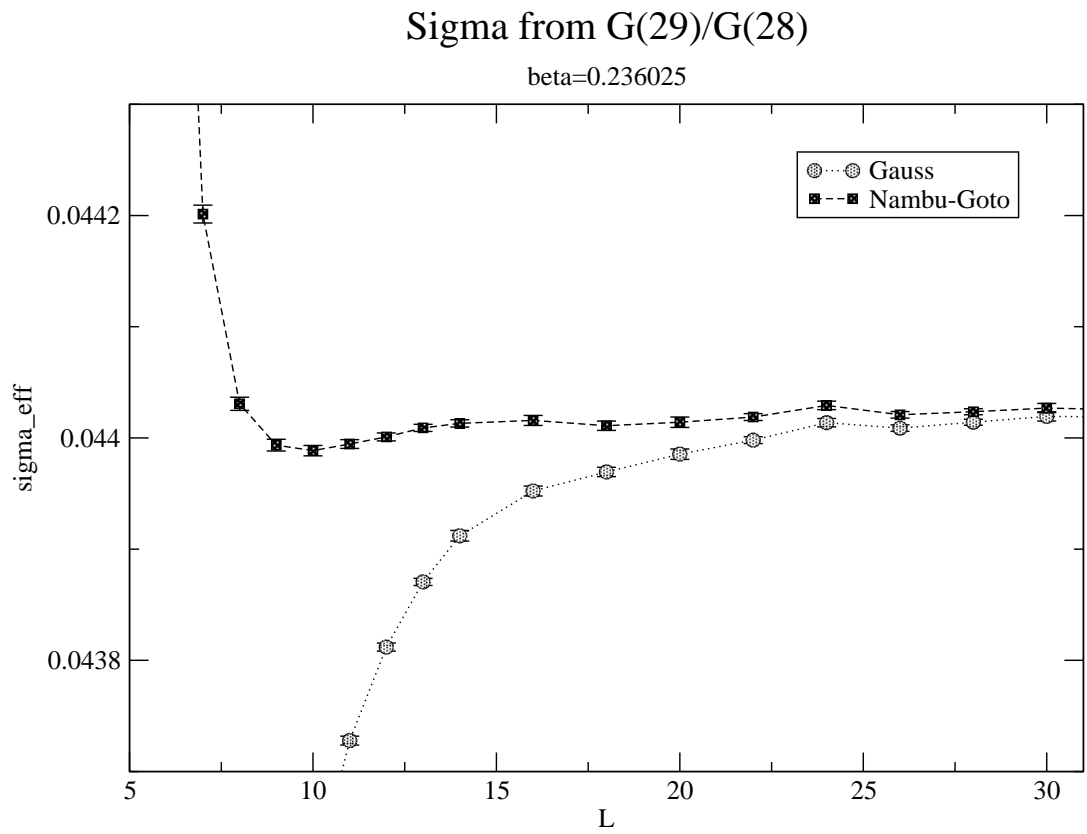}
\end{center}
\caption{
\label{sigmaR}
Effective string tension computed from $G(29)/G(28)$ using
eq.~(\ref{partitionfunctionasaseries3d}) for $\beta=0.236025$.
For Nambu-Goto and free field string states. 
}
\end{figure}

Finally in figure~\ref{sigmaR} we show $\sigma_{\mbox{\tiny{eff}}}$
 obtained from 
$G(29)/G(28)$ at $\beta=0.236025$.  For this particular value of $r$ 
we have added smaller values of $L$: $L=8,9,10,11,12$ and $13$. 
In the case of the Nambu-Goto energy-levels, $\sigma_{\mbox{\tiny{eff}}}$ 
is almost constant
down to $L=14$, while for the free field energy levels there is a clear $L$
dependence in the same range of lattice sizes. 

\begin{figure}
\begin{center}
\includegraphics[width=12.3cm]{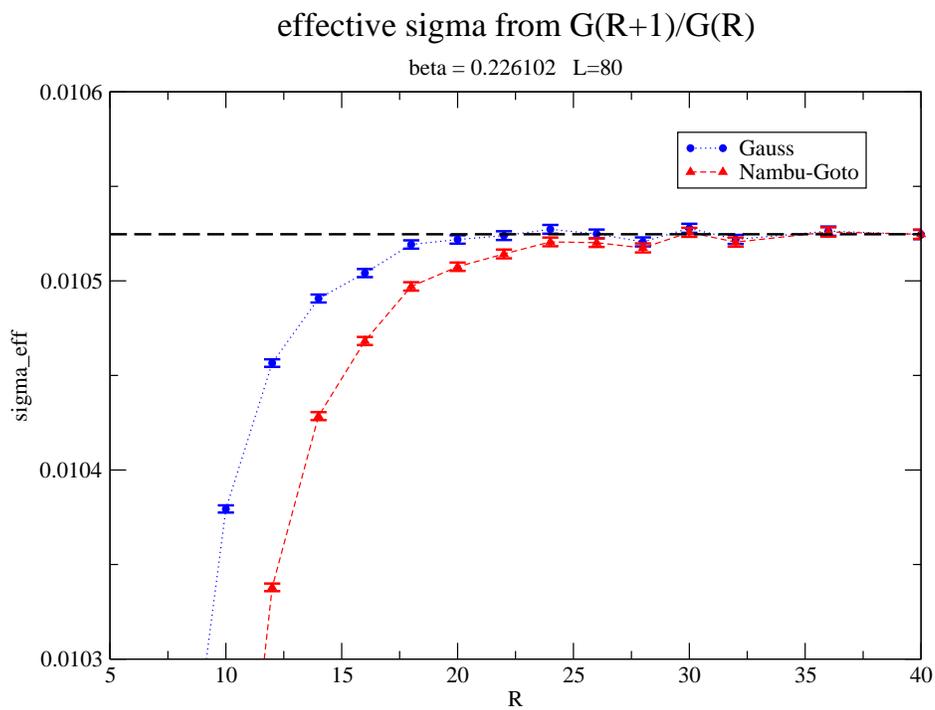}
\end{center}
\caption{
\label{sigmaold}
Effective string tension computed for $L=80$ using
eq.~(\ref{partitionfunctionasaseries3d}) for $\tilde \beta=0.226102$.
For Nambu-Goto and free field string states.
Data taken from table~3 
of~\cite{Caselle:2004jq}.
}
\end{figure}

In figure \ref{sigmaold} we have evaluated the effective string tension 
for $\tilde \beta=0.226102$ and $L=80$. The data are taken from table~3 
of~\cite{Caselle:2004jq}. Our estimate of~\cite{Caselle:2002ah, Caselle:2004jq}
for the string tension is $\sigma=0.010560(18)$ and $\xi=3.09(1)$ for the 
correlation length, \emph{i.e.}
$\sigma \xi^2 =0.1008(7)$. Similar to figure~\ref{Lmany4} for 
$\tilde \beta=0.236025$, we see for large $L$ that the Nambu-Goto
matches the data less good than the free string prediction. This fact 
had also been pointed out in~\cite{Caselle:2004jq}. 

We may summarize the results of this first part of our analysis in the following two points:
\begin{itemize}
\item
For all the values of $\beta$ that we studied the $L$ dependence of the interquark potential
is well described by the  Nambu-Goto effective string down to rather small values of $L$ (of the order of twice
the deconfinement length). In this respect the Nambu-Goto string behaves much better than the simple free string
model.
\item
String-related quantities (like the effective string corrections we are looking for) show much smaller scaling deviations than bulk observables. 
For instance, for the three values of $\beta$ that we studied,
due to the very small values of $\xi$ the universal combination
$\sigma\xi^2$ is still far from the continuum limit value, while in the 
same samples the $L$-dependence of the interquark potential is well described by the Nambu-Goto effective string (see the collapse of curves in figures
\ref{Lmany4} to \ref{Lmany2}), independently of the value of $\xi$.
\end{itemize}

\subsection{The ground state energy}
\label{spectrum}
The ground state energy can be easily determined since, for large $L$, it 
dominates the string partition function:
\eq
\label{atlargel}
Z_{\mbox{\tiny{spin}},Q\bar{Q}}(r,L) \simeq e^{-L E_0(r)} \;\;\;\;\;\; \mbox{for $L$ large}
\en

In this limit, it is easy to make connection with the previous discussion: 
\begin{equation}
\frac{G(r+1)}{G(r)}  \simeq e^{-L [E_0(r+1)-E_0(r)]} \;\;\;\;\;\; \mbox{for $L$ large}
\end{equation} 
and, trivially
\begin{equation}
\label{sumE}
 E_0(r)=\sum_{\tilde r=0}^{r-1} [E_0(\tilde r+1)-E_0(\tilde r)]
\end{equation}
Above we found for $\tilde \beta = 0.236025$ and $L=80$ 
(where eq.~(\ref{atlargel}) holds within our numerical precision for the 
whole range of $r$ considered)
that $\sigma_{\mbox{\tiny{eff}}}$ (computed with the Nambu-Goto or the free field ansatz)
for $r < 10$ is clearly smaller than the asymptotic estimate of $\sigma$. 
It follows that also $E_0(r+1)-E_0(r)$ is clearly smaller than the Nambu-Goto
and also the free field prediction, using the asymptotic estimate of $\sigma$. 
On the other hand,  $\sigma_{\mbox{\tiny{eff}}}$ is constant within error-bars for the 
free field ansatz for $r> 10$ and also for the Nambu-Goto ansatz starting 
from $r > 15$ (where the difference of the Nambu-Goto and the free field theory
prediction is smaller than the error of our Monte Carlo data).

It follows directly from eq.~(\ref{sumE}) that, due to the contributions
from $r<10$, also for large values of $r$, $E_0(r)$ deviates from the 
effective string prediction (at least) by a constant. 

\begin{figure}
\begin{center}
\includegraphics[width=12.3cm]{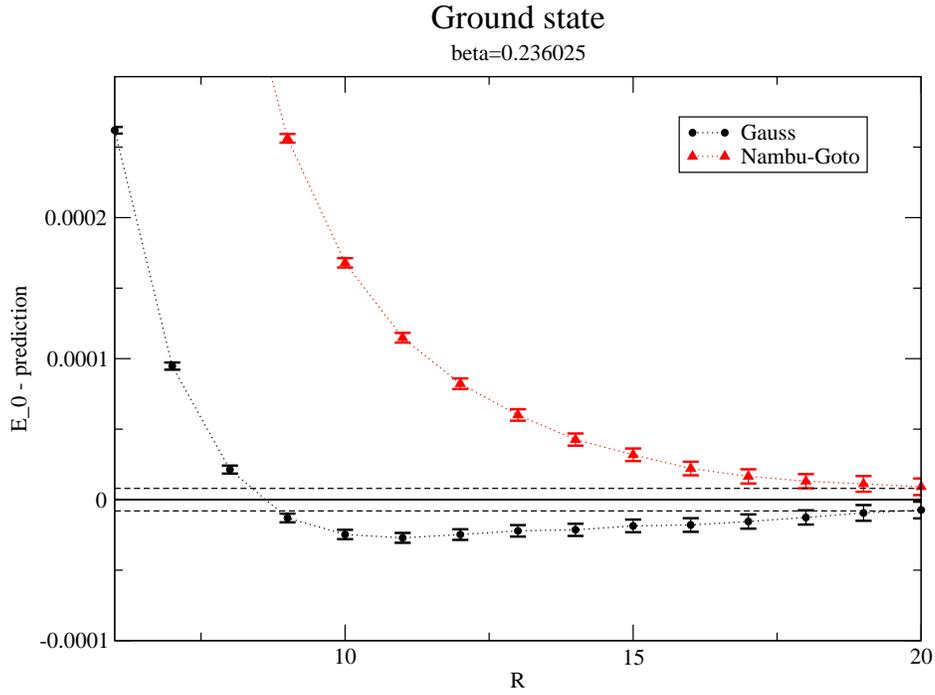}
\end{center}
\caption{
\label{energy0}
Ground state energy $E_0$ for $\tilde \beta = 0.236025$ taken from the fit
a.3 for $L_{\mbox{\tiny{min}}}=20$ discussed below. Note that $E_0$ is quite insensitive 
to the particular form of the fit and the value of $L_{\mbox{\tiny{min}}}$. The solid 
line indicates zero, the dashed lines indicate the error of $E_0(29)$ which
enters the constant that has been subtracted.  Note that the error in our 
estimate of the string tension $\sigma=0.044023(3)$ is not indicated in the 
figure.
}
\end{figure}

In figure~\ref{energy0} we have plotted $E_0$ for $\tilde \beta = 0.236025$ 
taken from the fit
a.3 for $L_{\mbox{\tiny{min}}}=20$ discussed below. Note that $E_0$ is quite insensitive 
to the particular form of the fit and the value of $L_{\mbox{\tiny{min}}}$. We give 
$E_0(r) - E_{0,\mbox{\tiny{prediction}}}(r)-const$. $const$ is chosen such that 
$E_0(29) - E_{0,\mbox{\tiny{prediction}}}(29)-const=0$. For the theoretical prediction, 
we have taken $\sigma=0.044023$ obtained above. For this value of $\sigma$
we get $const \approx 0.16684$, for both NG and free field theory. 
For Nambu-Goto  $E_0(r) - E_{0,\mbox{\tiny{NG}}}(r)-const$ is clearly larger 
than 0 for $r<18$. On the other hand, for the free field prediction, 
$E_0(r) - E_{0,\mbox{\tiny{free}}}(r)-const$ is slightly smaller than 0 for 
$9 < r < 20$, while it becomes positive for $r \le 9$. Here one should note 
that for $\sigma=0.0440245$ instead, $E_0(r) - E_{0,\mbox{\tiny{free}}}(r)-const$ is 
consistent with zero for all $r \ge 10$.

\subsection{The excited levels}
\label{excitedlevelsubsect}
From the free theory as well as from the Nambu-Goto effective string, 
we expect that the energy gaps $E_n-E_0$ are decreasing functions of 
the interquark distance $r$; in particular, for any $r_0$ and $n$ there 
exists an $r$ such that $E_n(r)-E_0(r)<E_1(r_0)-E_0(r_0)$. For such a 
choice of $r_0$, $r$ and $n$, we have:
\begin{eqnarray}
& & G(r)/G(r_0) = \exp \left(-[E_0(r)-E_0(r_0)] L\right) \;
\frac{\sum_{m=0}^{\infty} w_m \exp\left(-[E_m(r)-E_0(r)] L\right)}
     {\sum_{p=0}^{\infty} w_p \exp\left(-[E_p(r_0)-E_0(r_0)] L\right)} \nonumber \\ 
&=&
 \exp\left(-[E_0(r)-E_0(r_0)] L\right) \; \left[1 + \sum_{m=1}^{n} w_m \exp\left(-[E_m(r)-E_0(r)] L\right)
     + \dots \right]
\end{eqnarray}
that is, the nontrivial $L$-dependence of $G(r)/G(r_0)$ is dominated by the 
first $n$ states at the distance $r$, while the states at $r_0$ play virtually
no role. Hence, 
a good matching of $G(\tilde r+1)/G(\tilde r)$ with the NG  prediction
for $\tilde r > r_0$ implies that
also the first few energy gaps at $r$ have to follow the NG prediction, 
and vice versa. 

In the following, we determined the energy gaps from the correlation function $G(r)$. To this end, we performed a set of two- and three-parameter fits. Let us look at them in detail:

\begin{itemize}
\item two-parameter fits:

\begin{itemize}
\item[a.1] ``Two-state''  ansatz
\eq
\label{ia}
Z= e^{-E_0 L} + e^{-E_1 L} 
\en
Here, the fit parameters are $E_0$ and $E_1$. This ansatz makes no use of the string-theory.
\item[a.2] ``Free-string''  ansatz
\eq
\label{ib} 
Z= \sum_{n=0}^{\infty} P(n) e^{-E_n L}
\en
with $E_n=E_0 + n \Delta$, and $P(n)$ is the number of partitions of $n$; 
in this case, the fit parameters are $E_0$ and $\Delta$. Here we assume a string-spectrum 
as given by the free bosonic string. However, we allow $E_0$ and $\Delta$ to be independent. 

\item[a.3] ``Nambu-Goto'' ansatz
\eq
\label{ic}
Z= \sum_{n=0}^{\infty} P(n) e^{-E_n L}
\en
with $E_n= E_0 + \Delta_n$, and $P(n)$ is the number of partitions of $n$.
The energy-gaps $\Delta_n$ are given by eq.~\ref{enarvis}, where we allow for some $\sigma'$, 
not related to the $E_0$ in the ansatz, \emph{i.e.} the free parameters of the fit are $E_0$ and $\sigma'$.

\end{itemize}
\item three-parameter fits with:
\begin{itemize}
\item[b.1]  ``Three-state'' ansatz
\eq
\label{iia}
Z= e^{-E_0 L} + e^{-E_1 L} + 2e^{-E_2 L}
\en
The fitted parameters are $E_0$, $E_1$ and $E_2$. Here, we have used 
the degeneracy of the second excited state as theoretical input.
\item[b.2]  ``Modified free-string'' ansatz
\eq
\label{iib}
Z= \sum_{n=0}^{\infty} P(n) e^{-E_n L} 
\en
with: $E_0=a$, $E_1=a+b$, and $E_n=a+b+(n-1)c$ for $n>1$. 
In this case, the fitted parameters are $a$, $b$ and $c$.
As in the fit of type a.2 we assume the linear rising of the energy levels typical of the free string, but 
allow
the first energy gap (the parameter $b$ in the above equation) to behave independently 
of
the remaining energy
gaps.

\end{itemize}
\end{itemize}

The 
purpose of
this choice is 
to distinguish 
between the ``Nambu-Goto'' and the free string behaviour.

In the following, we report the results
for the sample at
$\tilde\beta=0.236025$ which, according to the 
findings discussed in
the previous section should be essentially unaffected by scaling violations as far as string-related quantities are concerned. 
In particular, we selected
the results of the fits at four values of the
interquark distance:  $r=4,9,19,29$. As mentioned above, for this value of 
$\tilde\beta$ 
we 
run simulations corresponding to
the following values of $L$:
$L=80,50,40,30,28,26,24,22,20,18,16,14,13,12,11,10,9,8$.
In the fits, only
the data for $L_{\mbox{\tiny{min}}}\geq14$ were included; then we studied the dependence of our results as $L_{\mbox{\tiny{min}}}$ was increased. Obviously, as $L$ is increased, 
the relative contribution of the ground state gets larger and larger,
and higher states in the spectrum become
negligible.

\begin{figure}
\centerline{\includegraphics[height=100mm]{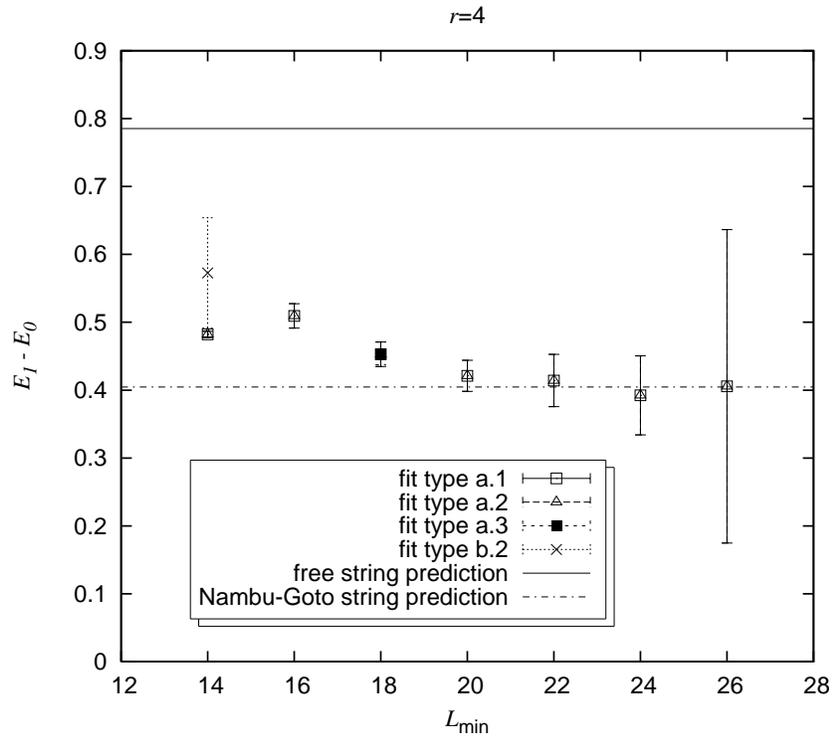}}
\vspace{1cm}
\caption{Comparison of the results for the first energy gap $E_1 - E_0$ obtained from different 
ans\"atze, as a function of the minimum value of $L$, at $\tilde{\beta}=0.236025$ for $r=4$.}
\label{r04deltaevslminfig}
\end{figure}

\begin{figure}
\centerline{\includegraphics[height=100mm]{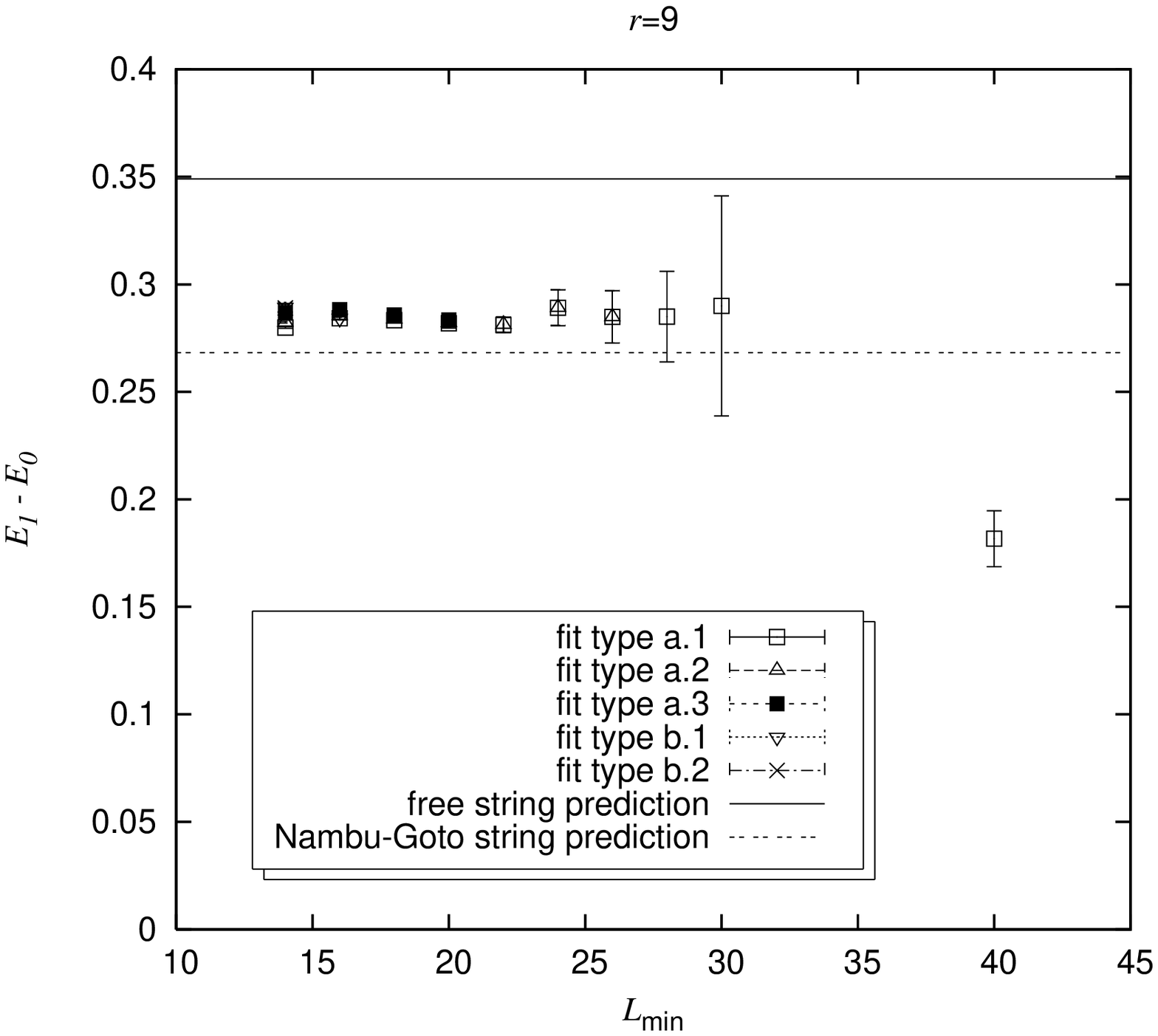}}
\vspace{1cm}
\caption{Same as in previous figure, but for $r=9$.}
\label{r09deltaevslminredchisqvslminfig}
\end{figure}

\begin{figure}
\centerline{\includegraphics[height=100mm]{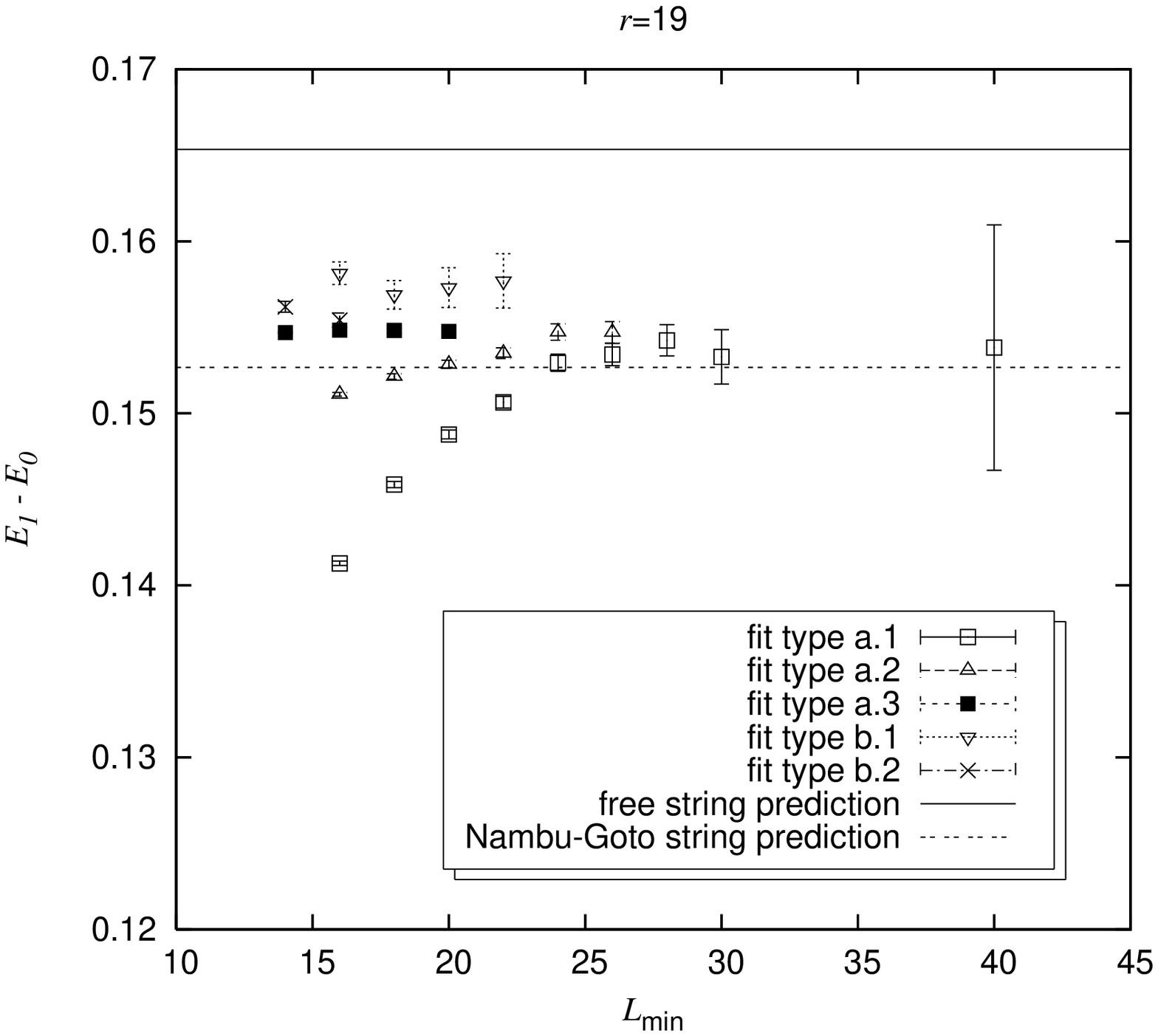}}
\vspace{1cm}
\caption{Same as in previous figures, but for $r=19$.}
\label{r19deltaevslminredchisqvslminfig}
\end{figure}

\begin{figure}
\centerline{\includegraphics[height=100mm]{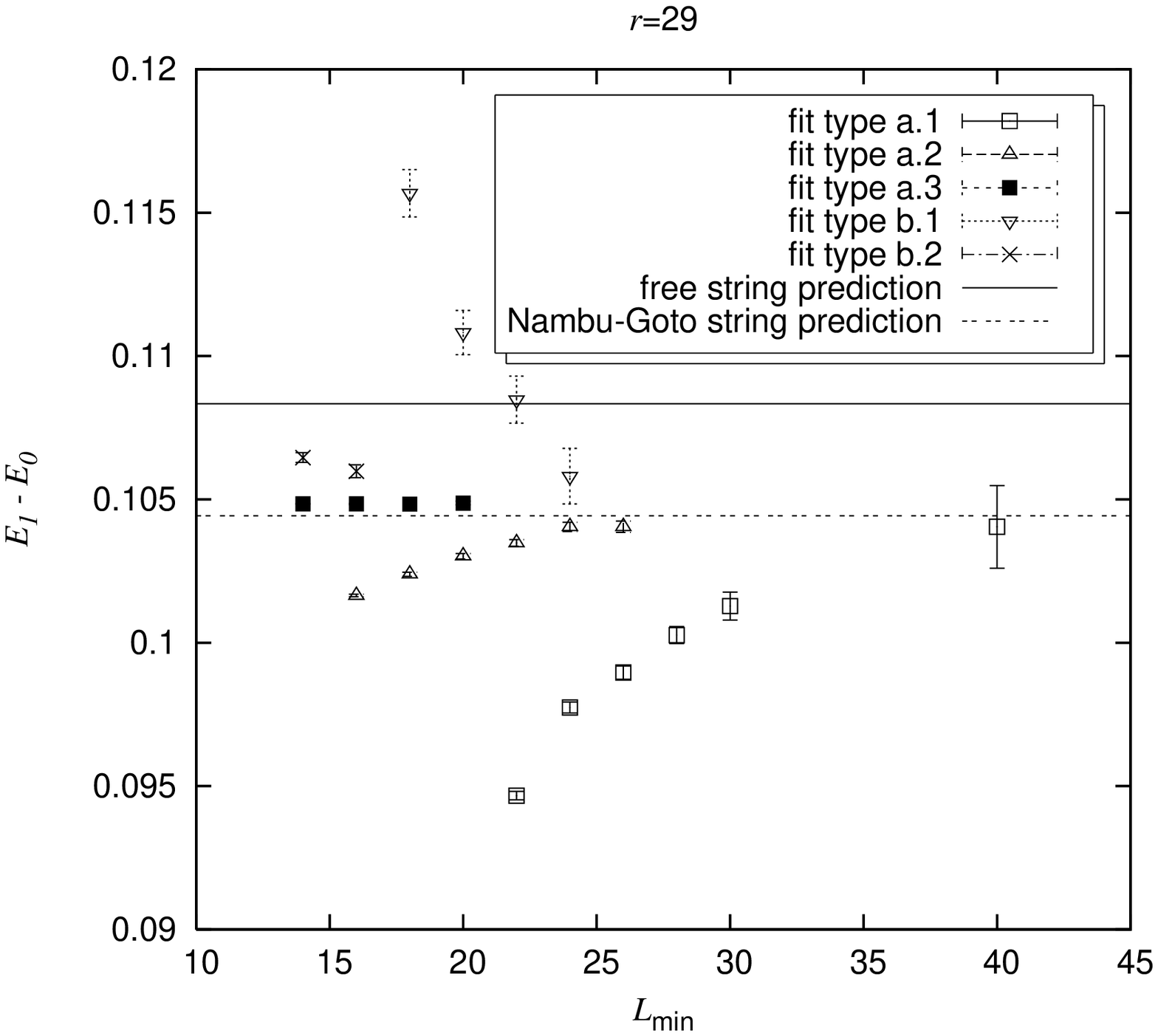}}
\vspace{1cm}
\caption{Same as in previous figures, but for $r=29$.}
\label{r29deltaevslminfig}
\end{figure}

In table~\ref{chisquaretab}, we report 
the 
reduced
$\chi^2$ of the fits and in 
figures~\ref{r04deltaevslminfig} to~\ref{r29deltaevslminfig} the best fit values for the lowest energy gap in the four cases.

Table~\ref{lightstatestab} displays the number of excited states, which, according to our results at various values of $r$, lie below the glueball mass threshold; note that at short distances the glueball threshold is close to the lightest $E_n$ states.

Some comments 
on these results are in order.
\begin{itemize}
\item
We expect that at large distance the Nambu-Goto string should describe
the interquark potential well. This is
indeed
clearly visible if we look at the $r=29$ entries in table~\ref{chisquaretab}: the fits of type a.3 (Nambu-Goto) 
show 
small reduced $\chi^2$ values already for
$L_{\mbox{\tiny{min}}}=14$, while fits of type a.1 and a.2 
display a bad behaviour until
large values of $L$ are reached, where  the
ground state dominates and only the first excited state gives
further significant contributions.  This is also clearly visible in
figure~\ref{r29deltaevslminfig} where the results for the energy gap 
of fit a.3 coincide with the Nambu-Goto prediction (the dashed
line) already for $L_{\mbox{\tiny{min}}}=14$,
while for the four other types of fit the data start 
for $L_{\mbox{\tiny{min}}}=14$ from values very
far apart and smoothly converge toward the Nambu-Goto limit as $L_{\mbox{\tiny{min}}}$ 
increases. Notice that the combination of
these two pieces of information 
(the reduced 
$\chi^2$ and 
the value of the 
$E_1-E_0$ gap as a function of $L_{\mbox{\tiny{min}}}$) tells us not only that
the lowest gap agrees numerically with the NG expectation, but also that the whole spectrum must be of the NG type. In particular one should note that the difference between the NG behaviour 
(a.1 fit) 
and  
the free string one
(a.2 fit)
can be clearly 
distinguished
by our data.

\item
A 
similar
pattern 
also occurs
for $r=19$, where, however, a systematic deviation with respect to the NG expectation
for the energy gap starts to be visible. The data smoothly converge towards 
a value of the $E_1-E_0$ 
gap which is
slightly larger than 
the
NG expectation, 
though 
still very far from the free string 
one 
(solid line).

\item 
For the two smallest values of $r$ the picture is completely different.
All the fits behave equivalently well and the 
reduced
$\chi^2$ cannot be used to 
distinguish among them. Also the best
fit values for the energy gap essentially coincide 
(this is particularly visible in the $r=9$ case where all the
symbols in 
figure~\ref{r09deltaevslminredchisqvslminfig} lie on top 
of 
each
other). This indicates that in this case the data are not precise enough 
to detect higher states in the spectrum and only $E_0$ and $E_1$ play a role
in the fits. This is not surprising, since, looking at eq.~(\ref{enarvis}), 
we see that as $r$ decreases the energy gaps become
larger and larger, making higher states negligible with respect to 
the two lowest ones.
As in the $r=19$ case, we see a clear disagreement with respect to 
the NG expectation, the observed energy gap
being larger than 
expected.

\end{itemize}

\begin{figure}
\centerline{\includegraphics[height=100mm]{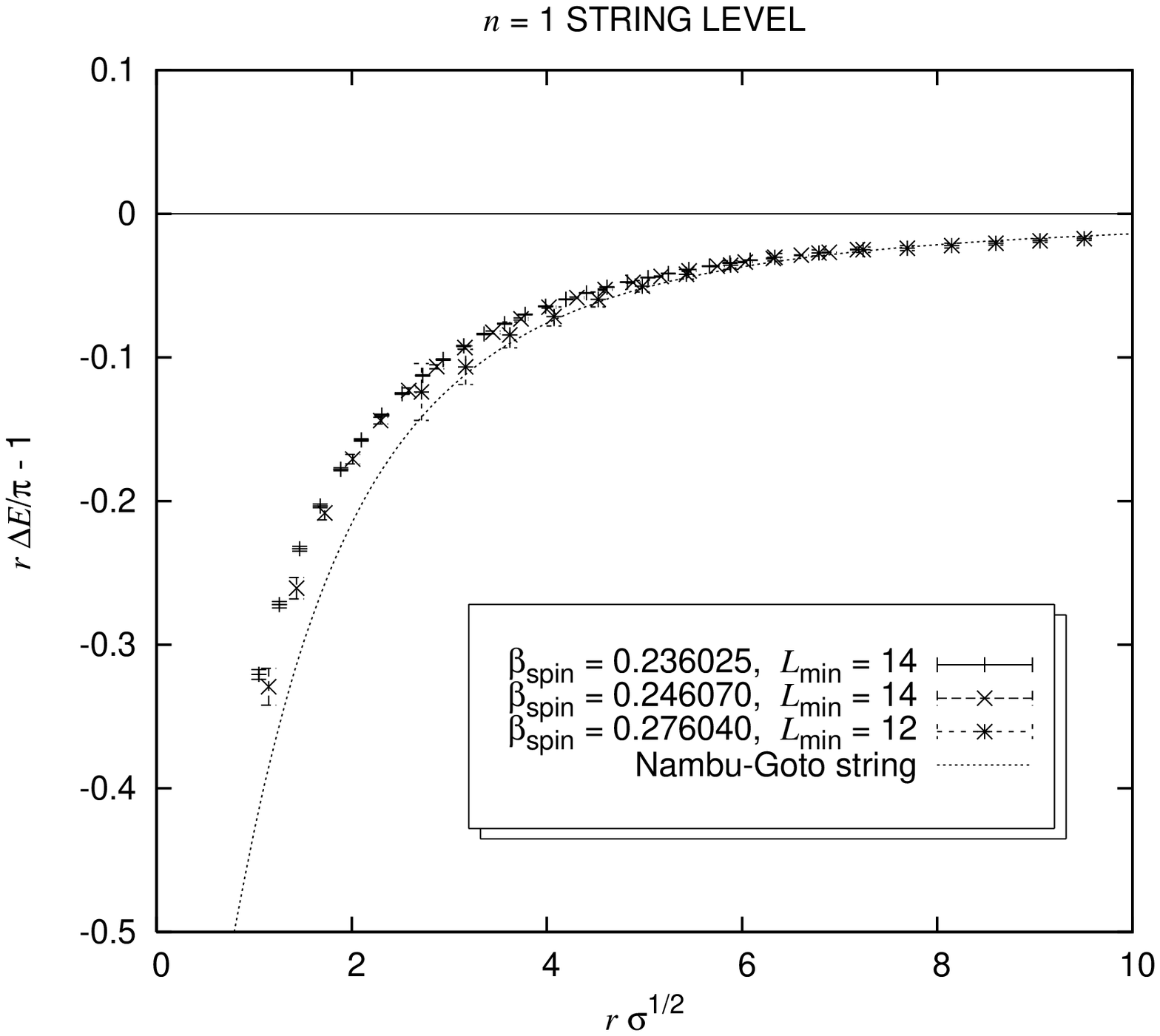}}
\vspace{1cm}
\caption{Relative deviation of the first energy gap $E_1 - E_0$, with respect to the free string prediction, 
as a function of the interquark distance; results are shown for different values of $\tilde{\beta}$, and are 
scaled in physical units. The format of this plot allows a direct comparison with figure~4 in 
\cite{Juge:2004xr}.}
\label{r04redchisqvslminfig}
\end{figure}

This disagreement is better appreciated in figure~\ref{r04redchisqvslminfig},
in which we plotted the relative deviation of the first energy gap $E_1 - E_0$ with respect to the free string prediction, as a function of the interquark distance. 
While the data are for all the values of $r$ very
far from the free string expectation (solid line in the figure), they definitely 
disagree from the NG
expectation for low values of $r$, and then nicely converge toward it as 
$r$ increases. The data for low values
of $r$ perfectly agree with those reported in \cite{Juge:2004xr}, where in 
fact a disagreement with respect to 
the NG picture was claimed. We confirm this disagreement at short distance, 
but being able to extend our
analysis to larger values of $r$ we can confirm (in agreement with our previous 
observations~\cite{Caselle:2005xy}) that at large distances the NG picture, both in the quantitative value of the energy gap $E_1-E_0$ and in the general pattern of the excited states, is fully restored.

Another interesting feature of 
figure~\ref{r04redchisqvslminfig}
is that it 
confirms the observation
we made in the previous section, about the fact that string-related 
quantities (like the energy gap $E_1-E_0$) 
appear not to be strongly affected by scaling violations. In the figure we plot the data obtained for 
the three values of
$\tilde\beta$
we studied, and --- except 
for small deviations in the $\tilde\beta=0.276040$
case --- the three samples
nicely agree in the whole range of values of $r\sqrt{\sigma}$.

\begin{table}[h]
\begin{center}
\begin{tabular}{|c|c|c|c|c|c|c|}
\hline
$r/a$ & $L_{\mbox{\tiny{min}}}/a$ & fit type a.1 & fit type a.2 & fit type a.3 & fit type b.1 & fit type b.2  \\
\hline
\hline  
 4 &  14 & \phantom{00}0.97 & \phantom{0}0.97 &           &                              &   0.83 \\
    &  16 & \phantom{00}0.68 & \phantom{0}0.68 &           &                              &           \\
    &  18 & \phantom{00}0.33 & \phantom{0}0.33 &  0.33  &                              &           \\
    &  20 & \phantom{00}0.23 & \phantom{0}0.23 &           &                              &           \\
    &  22 & \phantom{00}0.27 & \phantom{0}0.27 &           &                              &           \\
    &  24 & \phantom{00}0.31 & \phantom{0}0.31 &           &                              &           \\
    &  26 & \phantom{00}0.38 & \phantom{0}0.38 &           &                              &           \\
\hline
 9 &  14 &  \phantom{00}5.77  & \phantom{0}3.40 &  1.63 &  \phantom{0}1.84 &  1.73 \\
    &  16 &  \phantom{00}0.72  & \phantom{0}0.87 &  1.31 &  \phantom{0}0.81 &          \\
    &  18 &  \phantom{00}0.72  & \phantom{0}0.75 &  0.84 &                              &          \\
    &  20 &  \phantom{00}0.75  & \phantom{0}0.75 &  0.75 &                              &          \\
    &  22 &  \phantom{00}0.87  & \phantom{0}0.86 &          &                              &          \\
    &  24 &  \phantom{00}0.75  & \phantom{0}0.75 &          &                              &          \\
    &  26 &  \phantom{00}0.90  & \phantom{0}0.90 &          &                              &          \\
    &  28 &  \phantom{00}1.20  &                             &          &                              &          \\
    &  30 &  \phantom{00}1.80  &	                        &          &                              &          \\
    &  40 &  \phantom{00}0.25  &                             &          &                              &          \\
\hline
19 & 14 &                               &                              & 1.32   &                              & 1.43   \\
     & 16 &                  227.19  &                  13.35   & 0.90   & \phantom{0}1.26   & 0.87   \\
     & 18 & \phantom{0}57.56  & \phantom{0}5.31  & 1.01   & \phantom{0}0.74   &           \\
     & 20 & \phantom{0}19.02  & \phantom{0}3.45  & 1.14   & \phantom{0}0.82   &           \\
     & 22 & \phantom{00}8.20  & \phantom{0}2.42  &           & \phantom{0}0.95   &           \\
     & 24 & \phantom{00}0.82  & \phantom{0}0.35  &           &                              &           \\
     & 26 & \phantom{00}0.65  & \phantom{0}0.44  &           &                              &           \\
     & 28 & \phantom{00}0.29  &                              &           &                              &           \\
     & 30 & \phantom{00}0.18  &                              &           &                              &           \\
     & 40 & \phantom{00}0.36  &                              &           &                              &           \\
\hline
29 & 14 &                                 &                            &  1.21   &                              & 2.01  \\
     & 16 &                                 &              47.70     &  1.35   &                              & 0.92  \\  
     & 18 &                                 &              22.57     &  1.51   &                  15.19   &          \\
     & 20 &                                 &              10.84     &  1.59   & \phantom{0}5.58   &          \\
     & 22 &  171.38                    & \phantom{0}5.08 &            & \phantom{0}2.74  &           \\ 
     & 24 & \phantom{0}36.56    & \phantom{0}0.60 &            & \phantom{0}0.44  &          \\
     & 26 & \phantom{0}16.68    & \phantom{0}0.75 &            &                             &          \\
     & 28 & \phantom{00}3.92    &                             &            &                             &          \\
     & 30 & \phantom{00}2.39    &                             &            &                             &          \\
     & 40 & \phantom{00}0.05    &                             &            &                             &          \\
\hline
\end{tabular}
\end{center}
\caption{Some details about the analysis of data at $\tilde \beta =0.236025$: 
the table shows the values of $\chi^2_{\mbox{\tiny{red}}}$ as a function of the 
minimum lattice size in the time-like direction, from the fits corresponding to 
the various ans\"atze discussed in subsection~\ref{excitedlevelsubsect}, 
and at different values of $r$.}
\label{chisquaretab}
\end{table}
\begin{table}[h]
\begin{center}
\begin{tabular}{|c|c|}
\hline
$r/a$ & \# of states  \\
\hline
\hline
     4     &     1  \\
     9      &    3  \\
    14    &      4  \\
    19    &      5  \\
    24    &      6  \\
    29    &      7  \\
\hline
\end{tabular}
\end{center}
\caption{The number of excited $E_n$ states ($n\ge1$) lighter than the glueball threshold, as a function of the interquark distance (in lattice units); the table shows the results obtained from simulations at $\tilde \beta =0.236025$.}
\label{lightstatestab}
\end{table}

\section{Discussion and conclusions}
\label{conclusionssect}

The results of our analysis for the $\Z_2$ pure lattice gauge theory in $D=3$,
 which is expected to provide a 
prototypical model for quark confinement, show the following aspects.
\begin{enumerate}
\item A 
na\"{\i}ve 
description of the partition function associated with the 
confined $Q\bar{Q}$ sector of the 
theory expressed purely in terms of the standard effective string 
picture is not completely satisfactory, 
because the numerical results for the energy levels $E_n(r)$ do not 
agree with the corresponding theoretical 
spectrum, obtained by means of formal canonical quantisation of 
the Nambu-Goto string.

\item However, the observation that the differences $E_0(r+1)-E_0(r)$
are in agreement with the free string predictions  for
intermediate and large distances $r$, suggests that the 
deviations with respect to the string behaviour could be due to an 
overall shift in the spectrum
induced by extra terms, which are relevant at short length scales, 
but negligible in the infrared limit.

\item Our numerical data confirm the Nambu-Goto string prediction 
for the energy gap $E_1(r)-E_0(r)$ 
for large interquark distances $r$. 
In particular, there is a large range in $r$, where the 
precision of our numerical 
results allows to clearly distinguish between the Nambu-Goto 
and the free string prediction and where we see a better 
agreement with the Nambu-Goto string than with its free string approximation. 

\item 
Our fits indicate that also gaps $E_n(r)-E_0(r)$ with $n>1$ follow the 
Nambu-Goto prediction for sufficiently large $r$. Unfortunately, our data 
do not allow to make more quantitative statements on this issue.

\end{enumerate}
Furthermore, we have discussed how the results on the spectrum that we have 
summarized above are related with the behaviour of the ratio $G(r+1)/G(r)$
of Polyakov loop correlation functions that we have analysed in previous
studies \cite{Caselle:2002rm,Caselle:2002ah,Caselle:2004jq,Caselle:2004er,Caselle:2005xy}: A matching of $G(r+1)/G(r)$ with the Nambu-Goto string prediction
for large $r$ implies that also the gaps $E_n(r)-E_0(r)$ follow the 
Nambu-Goto string prediction.

It is interesting to compare these results with other studies 
that are available in the literature; in particular, in \cite{Juge:2002br,Juge:2003sz,Juge:2003ge,Juge:2004xr} the excitation spectrum was described in terms of the standard notation for the diatomic molecular 
physics, with states $\Gamma$ labelled by the component of the angular momentum of the gluon field along the interquark axis, and by the $CP$ 
eigenvalues.
The numerical method they used to extract the lowest energy in the various 
$\Gamma$ sectors goes through evaluation of  generalised Wilson loops,
whose spatial edges were replaced by well-suited combinations of paths, 
transforming according to the quantum numbers of the considered $\Gamma$ state. They focused their study on a regime of intermediate distances, and the results show rather good agreement with the predictions of the adiabatic approximation of the bag 
picture \cite{DeGrand:1975cf}, which provides a phenomenological model 
expected to interpolate between the short and long length scales; at large distances, however, the energies associated with the internal ``gluonic'' excitations of the stretched bag will become irrelevant to the lowest spectrum states, and a convenient description for the flux-tube will be given in terms of an effective string model. They also observe fine-structure deviations with respect to the $\frac{n\pi}{r}$ gaps expected from free-string model, that they compare with the Nambu-Goto spectrum (leaving room for the possibility that further physical effects can enter the effective string description in this regime).

In \cite{Majumdar:2004qx} a similar study was presented for the $SU(2)$ 
gauge model in $D=3$,
concluding that the energy gaps at finite lattice spacing
appear to be  well modeled by the Nambu-Goto prediction,
while the continuum extrapolation seems to favour the free
string description.

As mentioned above, our results suggest instead that in the regime of
intermediate to long distances the Nambu-Goto model 
(rather than the free-string one) correctly describes the spectrum 
of the theory, modulo an overall shift. As it concerns the 
interpretation of this shift, it is interesting to note that 
our analysis in the present work shows that  major 
contributions to this effect come at distances smaller 
than the inverse of the mass of the lightest glueball: therefore, the shift 
in the energy levels likely depends on the gauge model
which is considered and cannot be predicted in a pure string scenario.
On the other hand, the fact that the differences
$E_0(r+1)-E_0(r)$ appear to be better described by the
free string than the Nambu-Goto one in a wide range of intermediate 
distances might be
a (model-independent) effect due to the Liouville mode; 
the latter appears in the derivation of the Nambu-Goto effective
model from the reparametrization invariant string theory in
the framework of
covariant quantization, and, due to its intractability, it must be 
neglected in the calculations yielding the effective string 
spectrum~\cite{Billo:2005iv}. Neglecting the Liouville mode is allowed
 only in the critical dimension ($D=26$ for the Nambu-Goto string),
 in which it decouples from the theory.
In the present $D=3$ case the field should not be neglected, however one
 can argue that the effect of this approximation becomes less and less 
important as the 
interquark distance increases\footnote{In fact in~\cite{Billo:2005iv} 
it was shown that the spectrum obtained neglecting the Liouville mode 
exactly coincides with the Arvis one (obtained assuming the physics gauge),
 which is known to be correct at 
large interquark distances, since the anomalous term which breaks
the Lorentz invariance (the analogue of the Liouville field in this
framework) is known to decrease as a function of the interquark 
distance~\cite{Olesen:1985pv}.} --- a scenario which indeed appears to be compatible with our numerical results.

It would be very interesting to extend the present study to other models 
(especially to different values of $D$, which gives the coupling of the 
Liouville mode) to see if the present picture is confirmed, and to identify 
how the deviations from the Nambu-Goto effective string depend on the gauge 
group and on the number of space-time dimensions.

\vskip1.0cm {\bf
Acknowledgements.} This work was partially supported by the European Commission TMR programme HPRN-CT-2002-00325 (EUCLID). M.~P. acknowledges support received from Enterprise Ireland under the Basic Research Programme.

\end{document}